\documentclass[man]{apa6}

\usepackage[american]{babel}
\usepackage{csquotes}
\usepackage[style=apa,sortcites=true,sorting=nyt]{biblatex}
\usepackage{graphicx,epsfig}
\usepackage{amsmath}
\usepackage{amssymb}
\usepackage{color}
\usepackage[makeroom]{cancel}
\usepackage{verbatim}
\usepackage{bm}
\usepackage{hyperref}
\usepackage{fancybox}
\usepackage{color}
\usepackage{mdframed}
\usepackage{tikz}
\usepackage{caption}
\usepackage{subcaption}
\usepackage{threeparttable}
\usepackage{listings} %code
\usetikzlibrary{shapes.geometric, arrows} %flowchart
\usepackage{multirow}
\usepackage{threeparttable}
\usepackage{float} % figure in paper not after paper

\DeclareLanguageMapping{american}{american-apa}
\begin{document}
\title{A Comparative Evaluation of a Conditional Median-Based Bayesian Growth Curve Modeling Approach with Missing Data}
\shorttitle{GCM with Missing Data}
\threeauthors{Dandan Tang}{Xin Tong}{Jianhui Zhou}
\threeaffiliations{Department of Psychology, University of Virginia}{Department of Psychology, University of Virginia}{Department of Statistics, University of Virginia}
\authornote{This research was supported by the National Science Foundation under grant no. SES-1951038. Correspondence should be addressed to Dandan Tang. Email: gux8df@virginia.edu.}
\abstract{Longitudinal data are essential for studying within-subject change and between-subject differences in change. However, missing data, especially with nonnormal distributions, remains a significant challenge in longitudinal analysis. Full information maximum likelihood estimation (FIML) and a two-stage robust estimation (TSRE) are widely used for handling missing data, but their effectiveness may diminish with data skewness, high missingness rate, and nonignorable missingness. Recently,  a robust median-based Bayesian (RMB) approach for growth curve modeling (GCM) was proposed to offer a novel way to handle nonnormal longitudinal data, yet its effectiveness with missing data has not been fully investigated. This study fills this gap by using Monte Carlo simulations to systematically evaluate RMB’s performance relative to FIML and TSRE.  It is shown that, in general, RMB GCM is a reliable option for managing ignorable and nonignorable missing data in various data distributional scenarios. An empirical example further demonstrates the application of these methods. % and an \textbf{R} package is provided to facilitate the implementation of RMB GCM in real-world studies.  
} 

\maketitle
\section{Introduction}
Longitudinal data, characterized by repeated observations on the same subjects across multiple time points, are pivotal in studying within-subject change and between-subject differences in within-subject change.  Among the analytical techniques used for longitudinal data, growth curve modeling (GCM) stands out as a particularly effective technique due to its capability to delineate individual growth trajectory and evaluate variability among these trajectories (e.g., Curran, Obeidat, \& Losardo, 2010; Grimm, 2016).

While GCM has been widely used to analyze longitudinal data in social and behavioral sciences (e.g., Itzchakov et al., 2023; Parodi et al., 2025), an inevitable challenge for GCM is how to handle missing data. Data collection over long periods often results in participants missing certain measurement occasions, leading to incomplete datasets. For example, a meta-analysis on personality traits reported an average missing data rate of 44\% (Roberts, Walton, \& Viechtbauer, 2006).  Missing data, if not properly addressed, can result in inefficient and biased parameter estimates and lower statistical power, thereby impacting the reliability and validity of growth curve analyses (Ibrahim \& Molenberghs, 2009; Leeper \& Woolson, 1982).

To better understand data missingness, different missing data mechanisms have been proposed (Little \& Rubin, 2002). These mechanisms are categorized into three types: missing completely at random (MCAR), where the lack of data is independent of any observed or unobserved values; missing at random (MAR), where the missingness depends only on observed values and not on the values that are missing; and missing not at random (MNAR), where the missingness is related to the missing values themselves or unobserved latent variables. MCAR and MAR data are called ignorable missing data because they can be adequately addressed using standard methods such as maximum likelihood based techniques. If data are MNAR, explicit modeling of the missing data mechanism is required to avoid biased parameter estimation; thus, they are called non-ignorable missing data. Since MCAR data can be addressed effectively by most current missing data analytical methods, this paper focuses on handling MAR as an ignorable missing mechanism and MNAR as a nonignorable missing mechanism in analyzing practical longitudinal data in social and behavioral sciences, where data are rarely normal (Micceri, 1989). In this paper, we investigate the performance of GCM for nonnormally distributed data with missing values. 

Growth curve models can typically be estimated in the structural equation modeling (SEM) framework, where Full Information Maximum Likelihood (FIML) is widely applied to handle missing data due to its ability to deliver consistent and efficient parameter estimates with normally distributed data and ignorable data missingness (Enders \& Bandalos, 2001; Shin et al., 2017). However, when data distribution deviates from normality, FIML may lead to biased and inefficient estimates (Yuan \& Bentler, 2001). The two-stage robust estimation (TSRE) was introduced by Yuan and Zhang (2012) as an extension of FIML and a flexible approach for nonnormal data, effectively downweighting potential outliers (e.g., Cruz et al., 2023). Further research by Yuan, Tong, and Zhang (2015) has shown that TSRE produces less biased and more efficient parameter estimation than FIML for cross-sectional, nonnormal data modeling under both MAR and MNAR mechanisms. Recent findings by Tang and Tong (2024) highlighted that TSRE outperformed FIML for analyzing longitudinal nonnormal data; however, although both of them provided more accurate parameter estimates than machine learning based analytical techniques for nonignorable missing data, neither FIML nor TSRE yielded satisfactory estimation results, especially when distributions were nonnormal and missingness rates were high. Thus, dealing with longitudinal nonignorable missing data remains a challenge, and statistical techniques are desired to effectively handle this type of missingness in nonnormal longitudinal data.

An alternative way to address nonnormal data in GCM is using robust Bayesian estimation methods. Tong et al. (2021) introduced a robust Bayesian approach based on conditional medians, which is less sensitive to outlying observations and can yield more accurate and efficient estimates than traditional GCM based on conditional means. Although Bayesian estimation offers a flexible way to handle missing data (Linero \& Daniels, 2018), no study has investigated the performance of the robust median-based Bayesian GCM with missing values. To fill this gap and explore the effectiveness of the median-based method compared to other widely used methods in addressing ignorable and nonignorable missing data, this paper aims to systematically assess and compare the performance of FIML, TSRE, and the robust median-based Bayesian approach (RMB), and offer practical guidelines on method selection for substantive researchers conducting longitudinal studies. 

To make this paper self-contained, the next sections will briefly review FIML and TSRE, and introduce RMB with missing data analytical techniques in growth curve modeling. A Monte Carlo simulation will be presented to compare the performance of RMB against FIML and TSRE across various data distributions and missing data patterns. Subsequently, an empirical example using data from the National Longitudinal Survey of Youth 1997 Cohort (Bureau of Labor Statistics, U.S. Department of Labor, 2005) will demonstrate the application of the three estimation approaches. In the end of the paper, we will discuss the research outcomes and offer practical recommendations for substantive researchers.

\section{Growth Curve Modeling with FIML, TSRE, and RMB} 
\subsection{A brief review of growth curve models} 
In a longitudinal study involving a cohort of $N$ subjects, the $i$th subject is observed at $T_i$ $(i=1,2,...,N)$ different measurement occasions. Without loss of generality, we assume that all subjects are measured at a common set of $T$ time points. For the $i$th subject, let $\textbf{y}_i = (y_{i1}, ..., y_{iT} )$ represent the vector of observed scores, with $y_{it}$ denoting the score on the $t$th time point ($t= 1, ..., T$). An unconditional growth curve model is typically expressed as follows:
\begin{equation}
 \begin{aligned}
 & \textbf{y}_i = \mathbf{\Lambda}\textbf{b}_i + \textbf{e}_{i}, \notag	\\
 & \textbf{b}_i = \mathbf{\beta} + \textbf{u}_i, \notag
 \end{aligned}
\end{equation}
where $\mathbf{\Lambda}$ denotes a $T \times q$ factor loading matrix delineating the trajectory shapes of growth, the $q \times 1$ vector $\textbf{b}_i$ denotes the subject-specific random effects, and $\textbf{e}_{i}$ denotes the within-subject measurement errors. The random effects $\textbf{b}_i$ indicate variability between subjects, and their mean, $\mathbf{\beta}$, corresponds to the fixed effects applicable to the entire population. The residual vector $\textbf{u}_i$ is representative of the random components of $\textbf{b}_i$.

In traditional growth curve models, it is typically assumed that both $\textbf{e}_{i}$ and $\textbf{u}_i$ follow multivariate normal (MVN) distributions, i.e.,
\begin{equation}
 \begin{aligned}
 & \textbf{e}_{i} \sim \mathcal{MVN}_T(\textbf{0},\mathbf{\Phi}),\notag \\
 & \textbf{u}_i \sim \mathcal{MVN}_q(\textbf{0},\mathbf{\Psi}).\notag
 \end{aligned}
\end{equation}
In the above formulations, the subscript in the MVN distribution denotes the dimensionality of the random vector. The matrix $\mathbf{\Phi}$, a $T \times T$ covariance matrix for $\textbf{e}_i$, is often hypothesized to be diagonal ($\mathbf{\Phi} = \sigma^2_e\mathbf{I}$), implying that within-subject measurement errors are characterized by equal variances and are independent over time. 

\subsection{Full Information Maximum Likelihood (FIML)}
When missing data are present, FIML, also known as normal-distribution-based maximum likelihood, is one of the most frequently used techniques for growth curve model estimation. This method has been implemented in all popular SEM software like Mplus and lavaan (Muthen \& Muthen, 2007; Rosseel, 2012). Grounded in the principles of maximum likelihood estimation, FIML constructs a likelihood function for each case in the dataset, assessing the discrepancy between observed data and model predictions while utilizing only the available observations for each individual. It then maximizes the overall log-likelihood function by summing the individual log-likelihoods (Raykov, 2005; Schafer \& Graham, 2002; Shin et al., 2017).

Mathematically, for a given dataset with missing values, FIML estimates the parameters \( \bm{\theta} \) by maximizing the log-likelihood function:
\[
L(\bm{\theta}) = \prod_{i=1}^{n} L_i(\bm{\theta}),
\]
where \( L_i(\bm{\theta}) \) represents the log-likelihood contribution of the \( i \)th subject based on the observed data:
\[
L_i(\bm\theta) = \frac{1}{(2\pi)^{T/2} |\bm{\Sigma}_i^{obs}(\bm{\theta})|^{1/2}} 
\exp \left( -\frac{1}{2} (\mathbf{y}_i^{obs} - \bm{\mu}_i^{obs}(\bm{\theta)})' 
(\bm{\Sigma}_i^{obs}(\bm{\theta}))^{-1} (\mathbf{y}_i^{obs} - \bm{\mu}_i^{obs}(\bm{\theta})) \right).
\]
Note that $\mathbf{y}_i^{obs}$ denotes the observed data in $\mathbf{y}_i$, and $\bm{\mu}_i^{obs}(\bm{\theta})$ and $\bm{\Sigma}_i^{obs}(\bm{\theta})$ represent the model-implied expected mean vector and expected covariance matrix of the observed data, respectively.

Rather than imputing missing values directly, FIML uses all available data to estimate model parameters and standard errors. But this method functions similarly to an imputation mechanism by adjusting the model parameters based on observed data to infer the location of missing data points (Enders, 2023). FIML typically yields more accurate parameter estimates than simpler methods, such as listwise deletion, pairwise deletion, and imputation based on similar response patterns (Enders \& Bandalos, 2001). However, the multivariate normality assumption of this method is a limitation. Practical data often violate this assumption, resulting in biased parameter estimates.

\subsection{Two-Stage Robust Estimation (TSRE)}
TSRE (Yuan \& Zhang, 2011, 2012) is an advanced statistical technique to handle missing values and potential nonnormally distributed data within the SEM framework, and provides robust and reliable parameter estimates when the assumption of multivariate normality is violated (Yuan et al., 2015).

TSRE operates in two stages. In the first stage, the objective is to obtain robust estimates for the saturated mean vector \( \bm{\mu} \) and the covariance matrix \( \mathbf{\Sigma} \) by downweighting potential outliers. This is achieved using robust M-estimators (Yuan \& Zhang, 2012).  In the second stage, the growth curve model is applied to the robust mean vector estimate \( \hat{\bm{\mu}} \) and the covariance matrix estimate \( \hat{\mathbf{\Sigma}} \) obtained from the first stage. The model parameters \( \bm{\theta} \), standard errors, and the associated test statistics are estimated by minimizing a fit function \( F \):
\[
\hat{\bm{\theta}} = \arg \min_{\bm{\theta}} F(\hat{\mu}, \hat{\Sigma}, \bm{\theta}),
\]
where the fit function \( F \) typically quantifies the discrepancy between the model-implied covariance matrix \( \mathbf{\Sigma}(\bm{\theta}) \) and the robust covariance matrix estimate \( \hat{\mathbf{\Sigma}} \), as well as the model-implied mean vector \( \bm{\mu}(\bm{\theta}) \) and the robust mean vector estimate \( \hat{\bm{\mu}} \).

By dividing the estimation process into two stages, TSRE can be viewed as an extension of FIML with the additional feature to handle nonnormal data. Moreover, when auxiliary variables (not included in the original GCM) are available to explain the data missingness, TSRE can flexibly incorporate those auxiliary variables in the first stage to convert nonignorable MNAR data to ignorable MAR data, enhancing the accuracy and reliability of GCM analyses in the presence of missing values (e.g., Tong et al., 2014).

\subsection{Robust Median-based Bayesian (RMB) GCM}

Traditional GCM focuses on the conditional means of the outcome variables as $E(\textbf{y}_i|\textbf{b}_i)=\mathbf{\Lambda}\textbf{b}_i$. However, conditional means are sensitive to outlying observations and may be influenced by missing data. Robust median-based growth curve modeling, which is based on conditional medians $\textbf{Q}_{0.5}(\textbf{y}_i|\textbf{b}_i)$, provides a valuable alternative modeling technique (Tong et al., 2021). 

The growth curve model for conditional medians can be expressed as
\begin{equation}
 \begin{aligned}
 & \textbf{y}_i = \mathbf{\Lambda}\textbf{b}_{i0.5} + \textbf{e}_{i}, ~\mbox{with}~ \textbf{Q}_{0.5}(\textbf{e}_{i}|\textbf{u}_i) = 0,	 \\
 & \textbf{b}_{i0.5} = \mathbf{\beta}_{0.5} + \textbf{u}_i, ~\mbox{with}~ \textbf{Q}_{0.5}(\mathbf{u}) = 0, \\
 \end{aligned}
\end{equation}
where $\mathbf{\beta}_{0.5}$ denotes the fixed effects for the change of conditional medians of $\textbf{y}_i$, and $\textbf{u}_i$ denotes the random effects.  For the median-based GCM, model parameters can be solved by minimizing the sum of absolute residuals (Tong et al., 2021), namely,
\begin{equation}
\hat{\mathbf{\beta}}_{0.5} = \mathop{arg \, min}\limits_{\mathbf{\beta}} \sum_{i=1}^{N}|\textbf{y}_i-\mathbf{\mu}_i| ,
\end{equation}
where $\mu_i = \mathbf{\Lambda}\mathbf{\beta}_{0.5} + \mathbf{\Lambda}\mathbf{u}_i$. Note that we cannot use maximum likelihood estimation because the absolute function is not differentiable at 0.  

Tong et al. (2021) proposed a Bayesian method for estimation and inference of this model, in which asymmetric Laplace distributions are strategically used to convert the problem of estimating the median-based growth curve model into a problem of obtaining the maximum likelihood estimator for a transformed model.  Suppose that $y_i$ follows an asymmetric Laplace (AL) distribution with the skewness parameter $\tau =0.5$, i.e., \( y_i \sim AL(\mu_i, \sigma, 0.5) \). The likelihood function for \( N \) observations is 

\[
L(\theta|y_1, \ldots, y_N) \sim \frac{0.5^{2N}}{\sigma^N} \exp \left\{ - \frac{1}{\sigma} \sum_{i=1}^{N} |y_i - \mu_i| \right\}, \notag
\]
where \( \mu_i \) is a function of the model parameters \( \theta \) and  \( \sigma \) is a nuisance parameter. It is obvious that maximizing the likelihood function is equivalent to minimizing $\sum_{i=1}^{N} |y_i - \mu_i|$. 

Linking this property with Equation (2), we transform the median-based growth curve model in Equation (1) into 

\begin{equation}
\begin{aligned}
& y_{ij} \sim AL(\mu_{ij}, \sigma, 0.5),\notag \\
& \mu_{ij} = \mathbf{\Lambda}_j' \beta_{0.5} + \mathbf{\Lambda}_j' u_{ij},\notag \\
& u_i \sim f_u(0, \Psi_{0.5}), \notag
\end{aligned}
\end{equation}
where \( f_u(0, \Psi_{0.5}) \) represents a multivariate distribution with mean 0 and covariance matrix \( \Psi_{0.5} \), and \( \mathbf{\Lambda}_j' \) represents a row vector (the \( j \)th row of \( \mathbf{\Lambda} \) ) that extracts the relevant components for the specific time point \( j \).  For simplicity, we let \( f_u(0, \Psi_{0.5}) \) be a multivariate normal distribution, i.e., \( u_i \sim \mathcal{MVN}(\textbf{0},\mathbf{\Psi}_{0.5}) \).

Bayesian methods incorporating data augmentation provide a powerful and flexible framework for fitting models under an AL distribution.  Following Kozumi and Kobayashi (2011), the AL distribution can be represented using two variables: an exponential random variable \(W\) and a normal random variable \(Z\). Specifically, when \(W \sim \text{Exp}(\sigma)\) and \(Z \sim \mathcal{N}(0,1)\),

\[
y = \mu + \zeta\,W + \eta\,Z \sqrt{\sigma W}
\]
follows \(\mathrm{AL}(\mu,\sigma,\tau)\), where \(\zeta = \frac{1 - 2\tau}{\tau(1 - \tau)}\) and \(\eta^2 = \frac{2}{\tau(1 - \tau)}\). For \(\tau = 0.5\), the parameters simplify to \(\zeta = 0\) and \(\eta^2 = 8\). This construction simplifies posterior sampling, allowing MCMC methods like Gibbs sampling to efficiently estimate model parameters.

Bayesian estimation can automatically handle ignorable missing data. When missing data are MCAR or MAR, multiple imputations are applied when missing outcomes are sampled from the conditional distribution of the outcome variable. If the missing data are nonignorable, a joint model such as a selection model is required (Diggle \& Kenward, 1994).  In a selection model, we can specify a distribution for the full data (both observed and unobserved) and a model for the missingness process itself. Our study uses the following selection model for its flexibility,
\begin{equation} \label{missing}
logit(\Pr(y_{it} \text{ is missing})) = logit(\Pr(R_{it} = 1)) = \alpha_{0} + \alpha_{1} y_{i(t-1)} + \alpha_{2} y_{it},
\end{equation}
where the missingness indicator \(R_{it} = 1\) if \(y_{it}\) is missing and \(R_{it} = 0\) otherwise. The intercept \(\alpha_0\) represents the baseline log-odds of missingness, \(\alpha_1\) captures how the previous outcome \(y_{i(t-1)}\) influences missingness at time \(t\), and \(\alpha_2\) reflects the impact of the current outcome \(y_{it}\) on its probability of being missing. Consequently, the probability of missingness at time \(t\) depends on both the prior and current outcome values.

As long as \(\alpha_2\neq 0\) in Equation (3), the missingness process depends on unobserved current outcomes, indicating MNAR. When \(\alpha_2=0\) and  \(\alpha_1\neq 0\), the corresponding missing mechanism is likely to be MAR. When \(\alpha_2=0\) and  \(\alpha_1 = 0\), it indicates that the missing data mechanism could be MCAR. This flexible framework can also be expanded to include other predictors (e.g., covariates, latent factors) to address various patterns of missingness.

\section{A Monte Carlo Simulation Study to Compare the Performance of FIML, TSRE, and RMB in GCM}
In this section, the performances of FIML, TSRE, and RMB in GCM are evaluated and compared using a Monte Carlo simulation study, with various potential influential factors, including sample size, missing data mechanism, missingness rate, and data distribution. 

\subsection{Simulation design}
Following  Tong, Zhang, \& Yuan (2014) in which TSRE was investigated, in this study, we generate data from a growth curve model with four measurement occasions ( $T$ = 4).  As Figure 1 shows, in the population model, the fixed effects of the latent intercept and slope are 6 and 2 ($\beta=(\beta_L, \beta_S)'= (6, 2)'$), respectively. The variance of the latent intercept was 1 ($\sigma_L^2$ = 1), the variance of the latent slope is 1 ($\sigma_S^2$ = 1), and the correlation between the latent intercept and slope is 0. The variance of the within-subject measurement errors was also set at 1 ($\sigma_e^2$ = 1). 
\centerline{[ Insert Figure 1 here ]}
\begin{figure}
	\centering
		\includegraphics[width=7.5cm,height=8cm]{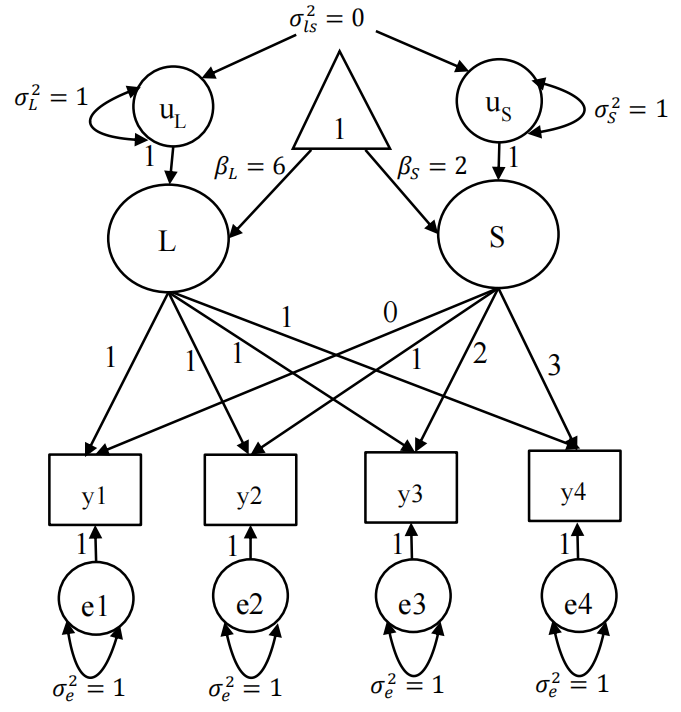}
	      \caption{Path diagram of a growth curve model.}  
       \label{fig:image1}
\end{figure}

As shown in Table 1, this simulation study manipulates four factors.  

1. The sample sizes considered are 100, 200, and 500, representing small, medium, and large sample sizes, respectively. These are typically used in growth curve modeling (e.g., Depaoli, Jia, \& Heo, 2023; Tong, Zhang, \& Zhou, 2021).  
\centerline{[ Insert Table 1 here ]}
\begin{table}[ht]
	\centering
	\caption{The Levels of manipulated factors}
	\label{t1} 
	\begin{tabular}{rr}
		\hline
		manipulated factors& levels \\ 
		\hline
		sample size & 100, 200, and 500 \\ 
		missing data mechanism& MAR and MNAR\\ 
		Missingness rates ($R$)  &  0\%, 5\%, 15\%, and 30\% \\ 
           \multirow{2}{*}{data distribution}  &  normal distribution, Student's t distribution,  \\  
              &   normal with $5\%$ outliers, and lognormal distribution \\ 
            
		\hline
	\end{tabular}
\end{table}

2. Two missing data mechanisms, MAR as an ignorable missingness and MNAR as a nonignorable missingness, are investigated. Under the MAR condition, observations \(y_{i2}, y_{i3},\) and \(y_{i4}\) are set to be missing when \(y_{i1} > c_1\), where \(c_1\) is a cutoff determined by the intended missingness rate.  If \(y_{i2}\) is observed, then \(y_{i3}\) and \(y_{i4}\) are subsequently set to be missing when \(y_{i2} > c_2\), and so forth (Tong, Zhang, \& Yuan, 2014). Hence, all missing values are MAR. Under the MNAR condition, an auxiliary variable ($Aux_i$) is defined and generated as $Aux_i = r\cdot b_{is} + \epsilon_i$, where $\epsilon_i$ follows a standard normal distribution $\epsilon_i \sim \mathcal{N}{(0,1)}$, $b_{is}$ represents the random effects of the slope, and $r$ is the regression coefficient of the latent slope. This simulation sets $r=0.8$ to maintain a high correlation between the auxiliary variable and the latent slope. Missingness in $y_{it}$ occurs if $Aux_i$ exceeds the value at a certain percentile $p_t$ determined by the desired missingness rate. Note that if the auxiliary variable is included in the analysis, missing data can be explained by the model and are thus MAR.  To maintain consistency with the real data analysis in the next section, which did not include any additional variables, we chose not to use the auxiliary variable in our simulation analysis. Thus, the missing data are MNAR.

3. Missingness rates ($mr$) of 0\%, 5\%, 15\%, and 30\% are examined. The first level (0\%) serves as a baseline that contains no missing data. The latter three levels represent small, medium, and large rates of missingness, respectively (Tong et al., 2021; Lei \& Shiverdecker, 2020). To yield this desired missingness level, for the MAR condition, we set $c_t$ ($t = 1,..., T-1$) at the upper $(1-[2t/(T-1)]\cdot mr)$th percentile of $y_t$. For the MNAR condition, $p_t$ ($t = 2, ..., T$) is set at the lower $(1-[2(t-1)/(T-1)]\cdot mr)th$ percentiles of the normal distribution $Aux_i\sim \mathcal{N}{(r\cdot b_{is},r^2+1)}$.

4. The study examines the influence of different data distributions, including normal distribution, Student's t distribution, normal distributed data with $5\%$ outliers, and lognormal distribution. These reflect various forms of within-subject measurement errors $\textbf{e}_{i}$. Specifically, for the normal distribution, $\textbf{e}_{i}\sim N(0,\sigma^2_e)$; for t distribution, $\textbf{e}_{i}\sim t_{(5)}{(0,\sigma^2_e)}$; for normal distribution with outliers, $5\%$ of the data are outliers that originate from a normal distribution with the same variance but shifted means $\textbf{e}_{i}\sim N(5,\sigma^2_e)$; and for lognormal distribution, $\textbf{e}_{i}\sim LN_{(0,1)}(0,\sigma^2_e)$.  The last three distributions capture common nonnormal conditions encountered in psychological research, i.e., heavy tails, outliers, and skewness (Micceri, 1989).

In total, our simulation study includes 84 conditions, computed as 3 (sample sizes) $\times$ 2 (missing data mechanisms) $\times$ 3 (rates of missingness excluding 0\%) $\times$ 4 (data distributions) $+$ 3 (sample sizes) $\times$ 1 (0\% missing data rate) $\times$ 4 (data distributions). For each condition, 500 datasets were generated and subsequently analyzed using FIML, TSRE, and RMB GCM. The RMB GCM was conducted in the R package \textit{rjags} (Plummer, 2003), while FIML and TSRE were conducted in the R packages \textit{lavaan} (Rosseel, 2012) and \textit{rsem} (Yuan \& Zhang, 2012), respectively. All programming codes are available at our \href{https://github.com/DandanTang0/GCM-with-Missing-Data}{GitHub} site.

For RBM GCM,  the following priors were used: the latent intercept $\beta_L \sim N(0,10^3)$, the latent slope $\beta_S \sim N(0,10^3)$, $\mathbf{\Psi} \sim InvWishart(I_2,3)$, where $I_2$ is $2 \times 2$ identity matrix, and the variance of the within-subject measurement errors $\sigma^2_e \sim InvGamma(0.001,0.001)$. The number of total iterations for each chain was set to 60,000, with the first half of the iterations used for the burn-in period.  Chain convergence was evaluated using Geweke's convergence diagnostic (Geweke, 1991). Chains that exhibited Geweke’s statistic values between -1.96 and 1.96 were considered to have achieved convergence. Subsequent analyses were conducted using converged chains. For simplicity but without loss of generality, we use the model without the selection structure for MAR data and a model with the selection structure for MNAR data because using the selection model for MAR data provided similar results (Shi \& Tong, 2018).

\subsection{Evaluation criterion}
The performances of FIML, TSRE, and RMB GCM were evaluated in terms of the relative bias and mean squared error of the model parameter estimates. The absolute relative bias quantifies that the average estimate among 500 replicates differs from the true value of the parameter relative to the true value itself. It can be expressed by $RB= 100\%\times \left| 
 \frac{\bar{\theta}-\theta}{\theta}\right|$, where $\theta$ denotes a population parameter,  and $\bar{\theta} = \frac{1}{500}\sum_{j=1}^{500}\hat{\theta}_j$, with $\hat{\theta}_j$  denoting a parameter estimate from the  $jth$ simulation replication,  $j = 1,...., 500$. When the population parameter is zero ($\theta = 0$), the ARB can be written as $RB= \left|\bar{\theta}-\theta \right|$. In general,  $RB < 10\%$ is considered an acceptable bias, and $RB < 5\%$ is considered as unbiased (Hoogland  \& Boomsma, 1998). 

The mean squared error measures the average squared difference between the estimated and actual values, which can be expressed as $MSE = \frac{1}{500}\sum_{j=1}^{500}(\hat{\theta}-\theta)^2$. MSE considers both the bias and the variance of the estimator, indicating how far the average estimate is from the true value as well as how spread out the estimates are. A lower MSE indicates a more accurate and precise estimator.

\subsection{Results}
Figures 2 to 5 display the results for the estimated average of latent slopes ($\beta_S$), which is often of primary interest for substantive researchers. The results for other parameters show similar patterns in terms of the comparison among the three methods and are thus included in the supplementary documents on our \href{https://github.com/DandanTang0/GCM-with-Missing-Data}{GitHub} site. 
\subsubsection{Methods comparison}
Under the MAR mechanism, FIML, TSRE, and RMB GCM had good and similar performances in terms of parameter estimation bias and mean squared error when data followed symmetric distributions, such as a normal distribution and a Student's t distribution.  For example, for normally distributed data, the relative bias for the estimated $\beta_S$ from these three methods was less than 5\%, as shown in Figure 2. When data followed nonsymmetric distributions, such as a normal distribution with $5\%$ outliers and a lognormal distribution, the three methods performed well at low to moderate missingness rates but RMB GCM performed better at high missingness rates. For instance, as shown in Figure 4, when the missing data rate was 15\%,  the relative bias of estimated $\beta_S$ based on these three methods was all smaller than 10\%, but when the missing data rate increased to 30\%, the relative bias from FIML and TSRE were above 10\% while the RMB GCM still yielded acceptable relative bias (i.e., < 10\%). 

Under the MNAR mechanism,  RMB GCM performed similar to FIML and outperformed TSRE with respect to parameter estimation bias and mean squared error when data followed symmetric distributions, such as a normal distribution and a Student's t distribution. For example, for the Student's t distribution, the relative bias for the estimated $\beta_S$ based on RMB GCM and FIML was almost the same regardless of the change of the missingness rate and sample size. When data followed nonsymmetric distributions, these three methods had different relative performances. Specifically, when data followed a normal distribution with $5\%$ outliers, RMB GCM performed better than FIML and TSRE in terms of estimation bias and efficiency, as shown in Figure 4. When data followed a lognormal distribution, the three methods performed similarly, as shown in Figure 5, and could not provide reliable parameter estimation when the missingness rate was above 15\%.

Detailed results are presented in the followings, regarding the manipulated factors in the simulation study. 

\subsubsection{Sample size}  
Under the MAR mechanism, FIML, TSRE, and RMB GCM showed stable estimation accuracy across different sample sizes for data following symmetric distributions. For example, for normally distributed data, the relative bias for estimated $\beta_S$ from these three methods was almost identical over varying sample sizes. For data following non-symmetric distributions,  the performance of FIML deteriorated slightly in estimating $\beta_S$ as the sample size increased, while RMB GCM and TSRE improved with larger sample sizes, which can be seen by comparing the subfigures in the second row of the figures in the top panel of Figures 4 and 5. 

Under the MNAR mechanism, the accuracy of FIML, TSRE, and RMB GCM improved in estimating $\beta_S$ with the increase of the sample size, as shown by comparing the subfigures in the first row of the figures in the bottom panel of Figures 2-5. 

\subsubsection{Rate of missingness} 
Under the MAR mechanism, FIML, TSRE, and RMB GCM exhibited stable estimation accuracy across different missingness rate conditions for symmetrically distributed data. For example, when data were normally distributed, the relative bias for estimated $\beta_S$ from these three methods was almost the same across the missingness rates. For the data following non-symmetric distributions, all three methods showed substantially decreased accuracy as the missingness rate increased.  For example, as shown in Figure 5, when the sample size was 100 and the missing data rate was 5\%, the relative bias of estimated $\beta_S$ was smaller than 5\%, but when the missing data rate increased to 30\%, the relative bias increased to more than 7\%.

Under the MNAR mechanism,  all three methods displayed declined performances with the increasing missingness rate, as shown by comparing the subfigures in the first row of the figures in the top panel of Figures 2-5. 

\subsubsection{Data distributions} 
Under the MAR mechanism, RMB GCM performed consistently well across all data distributions in estimating $\beta_S$. Both FIML and TSRE also performed well for symmetrically distributed data. For non-symmetrically distributed data, FIML and TSRE were comparable to RMB GCM at low to moderate missingness rates, and performed worse than RMB GCM when the missingness rate was high. This is illustrated in the first subfigure of the second row in the top panel of Figures 2 to 5.

Under the MNAR mechanism, RMB GCM again outperformed the other two methods across all data distributions in providing accurate and efficient model parameter estimates. FIML showed good performance across most data distributions except for normally distributed data with outliers. TSRE only showed its advantages for lognormally distributed data. These observations can be seen in the first subfigures of the first row in the top panel of Figures 2 to 5.

\centerline{[ Insert Figure 2 here ]}
\centerline{[ Insert Figure 3 here ]}
\centerline{[ Insert Figure 4 here ]}
\centerline{[ Insert Figure 5 here ]}

\begin{figure}[h]
    \centering
    \includegraphics[scale=0.2]{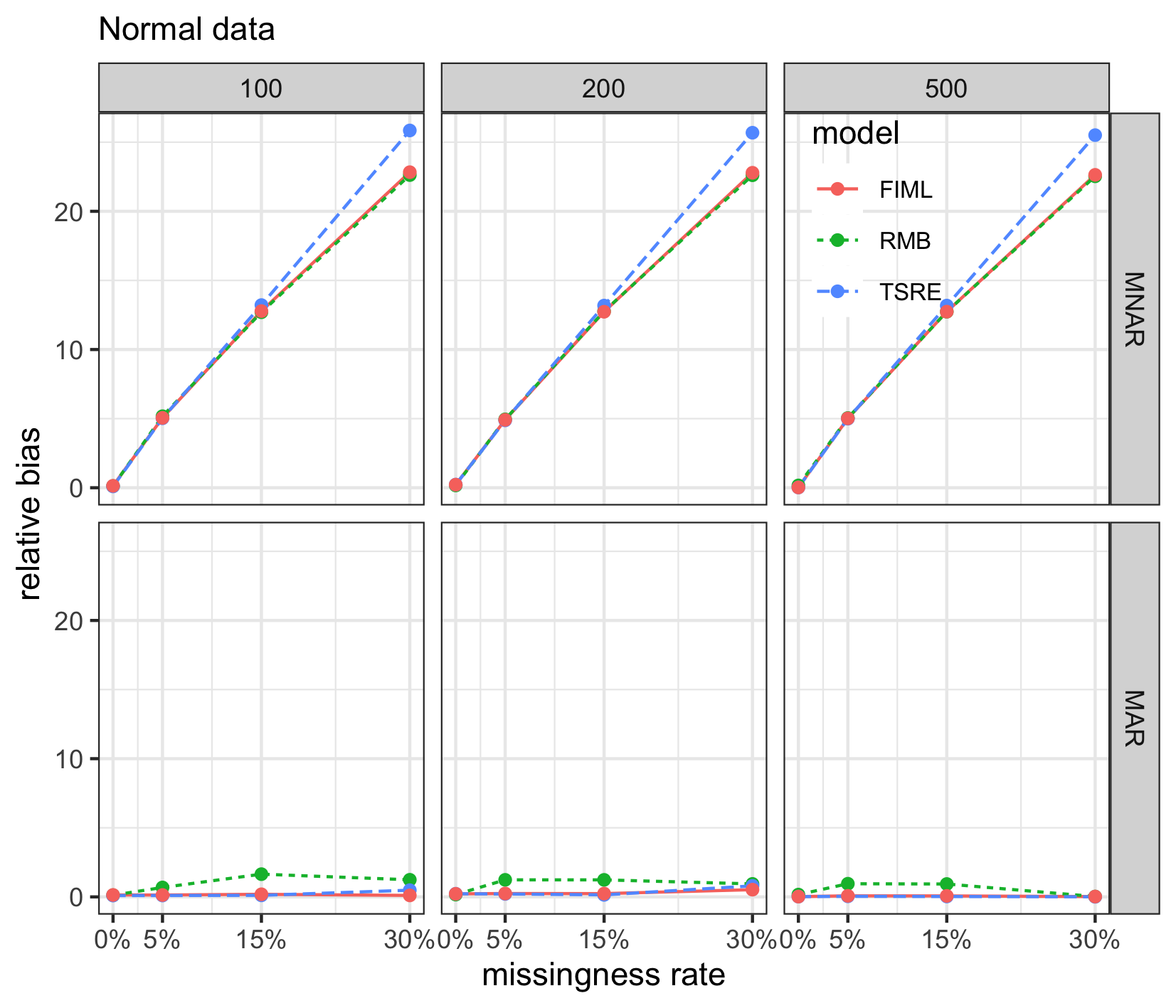} 
    
    \vspace{0.1cm} 
    
    \includegraphics[scale=0.2]{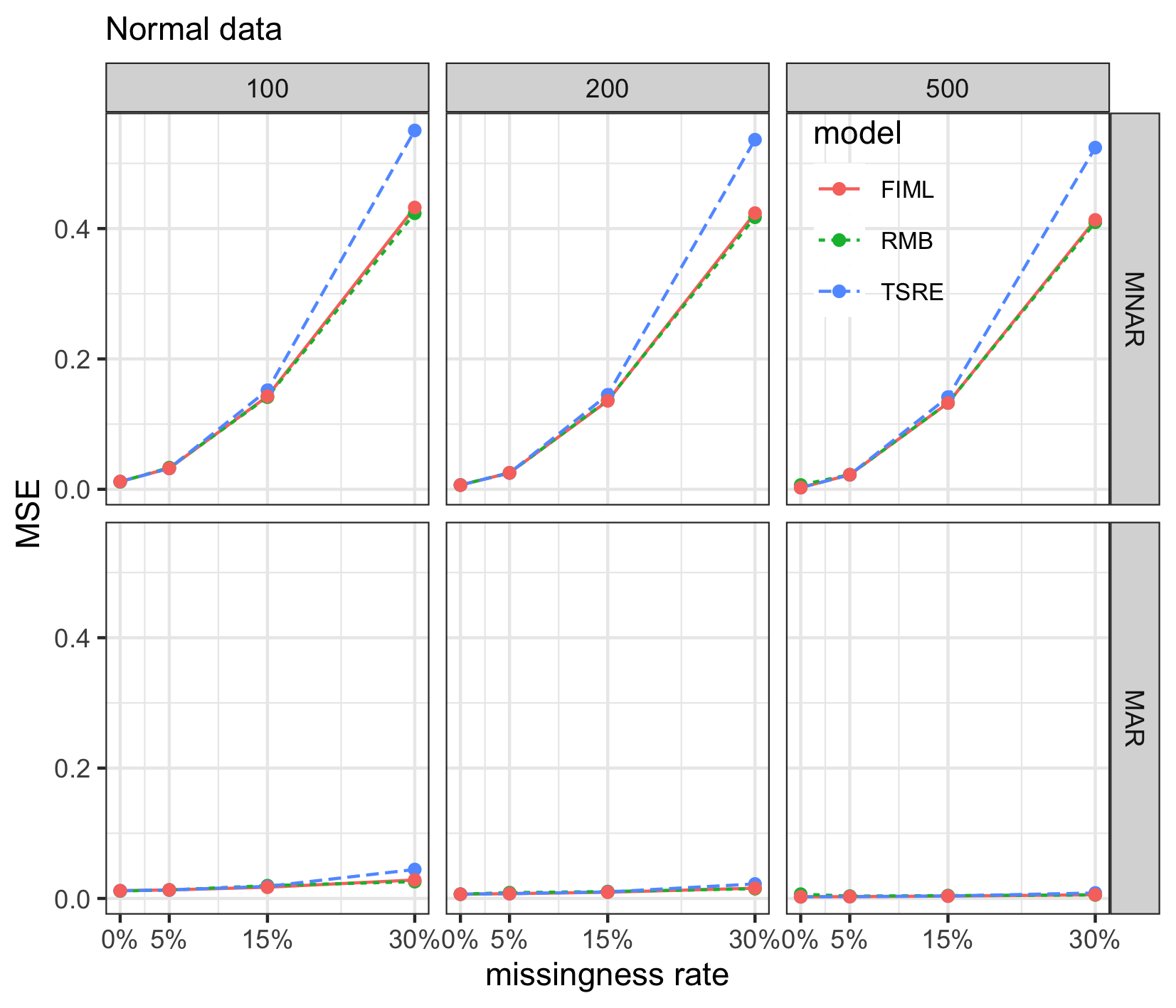} 
    \caption{The relative bias and mean squared error for the average of the latent slope estimates under the normal distribution} 
    \label{fig:image2}
\end{figure}

\begin{figure}[h]
    \centering
    \includegraphics[scale=0.2]{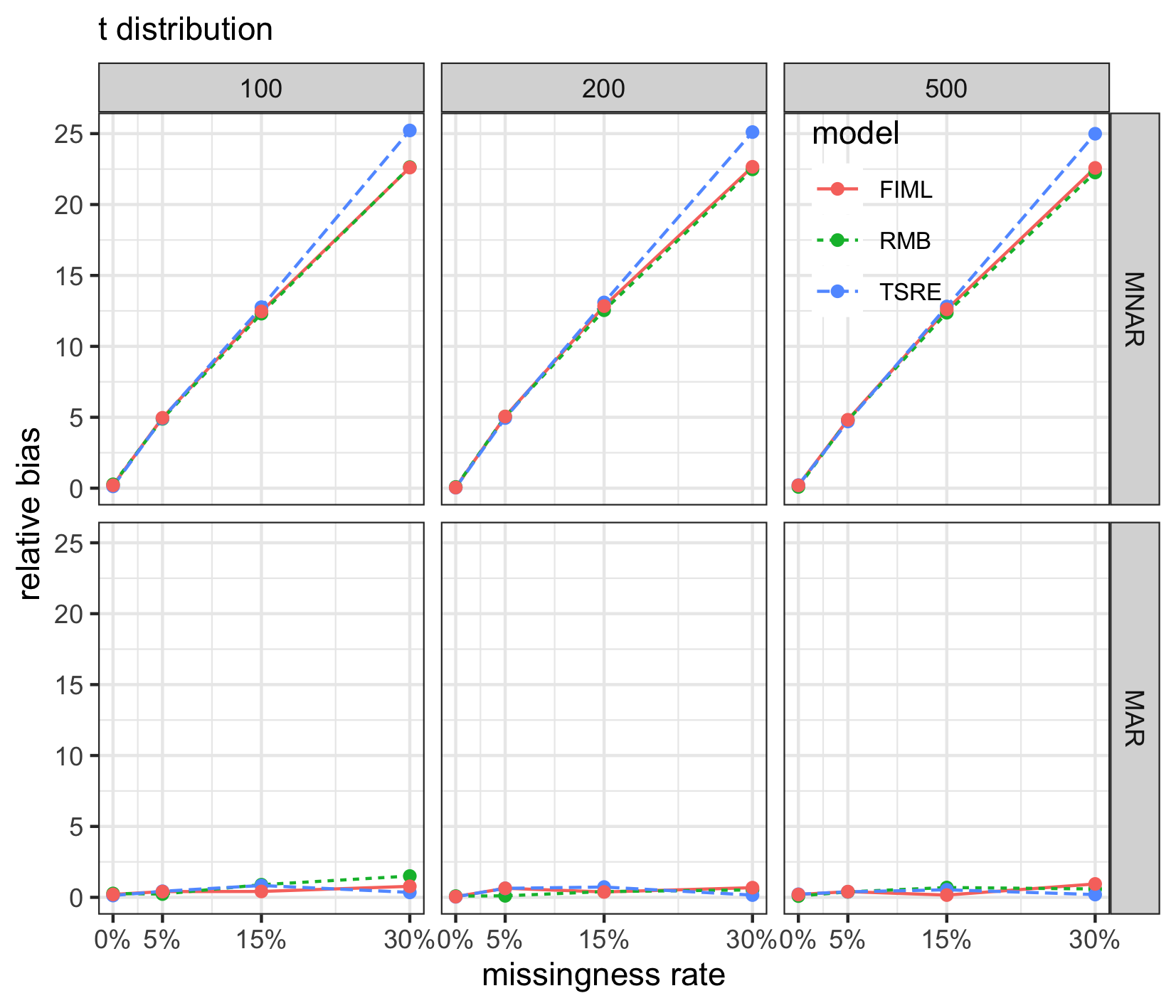} 
    
    \vspace{0.1cm} 
    
    \includegraphics[scale=0.2]{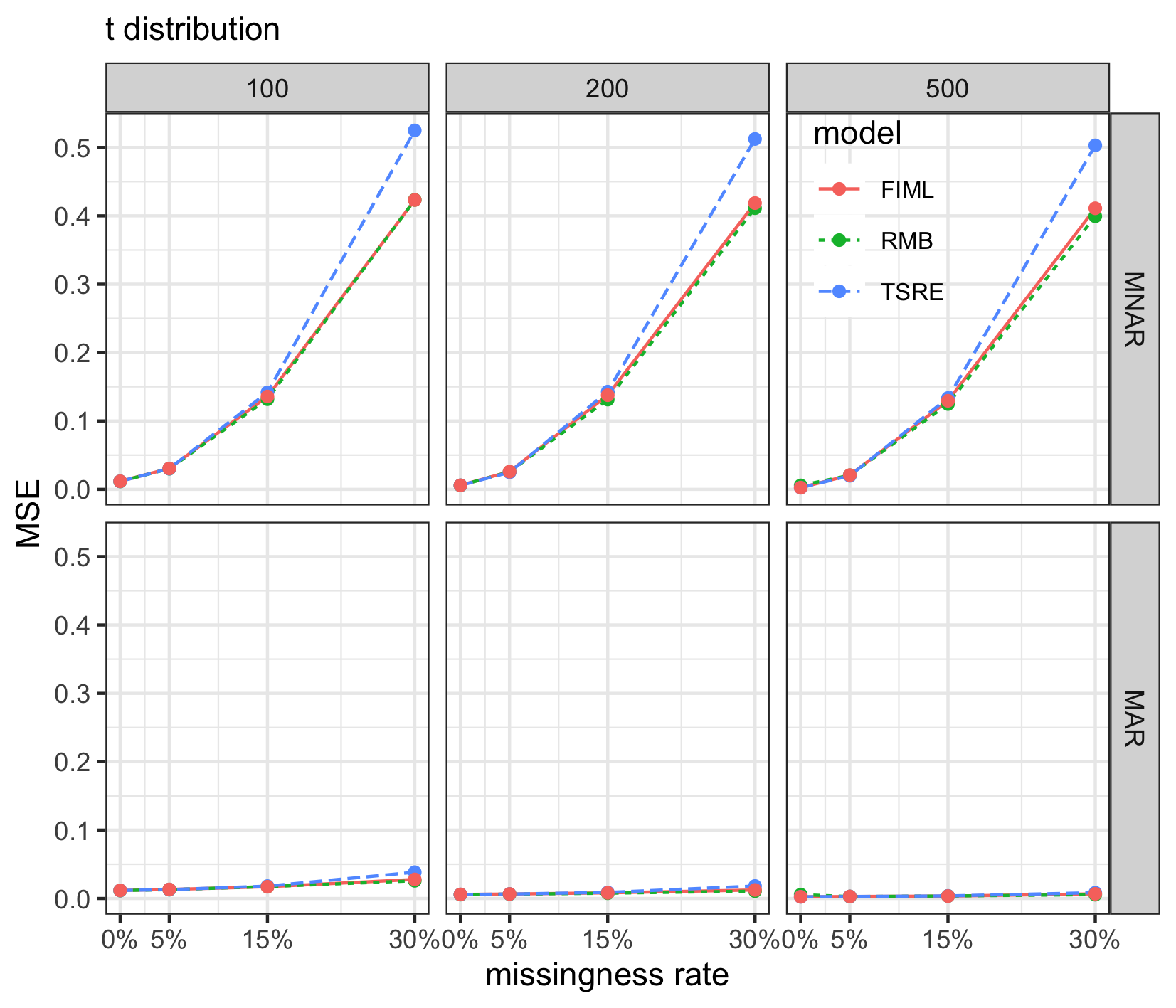} 
    \caption{The relative bias and mean squared error for the average of the latent slope estimates under the t-distribution} 
    \label{fig:image3}
\end{figure}

\begin{figure}[h]
    \centering
    \includegraphics[scale=0.2]{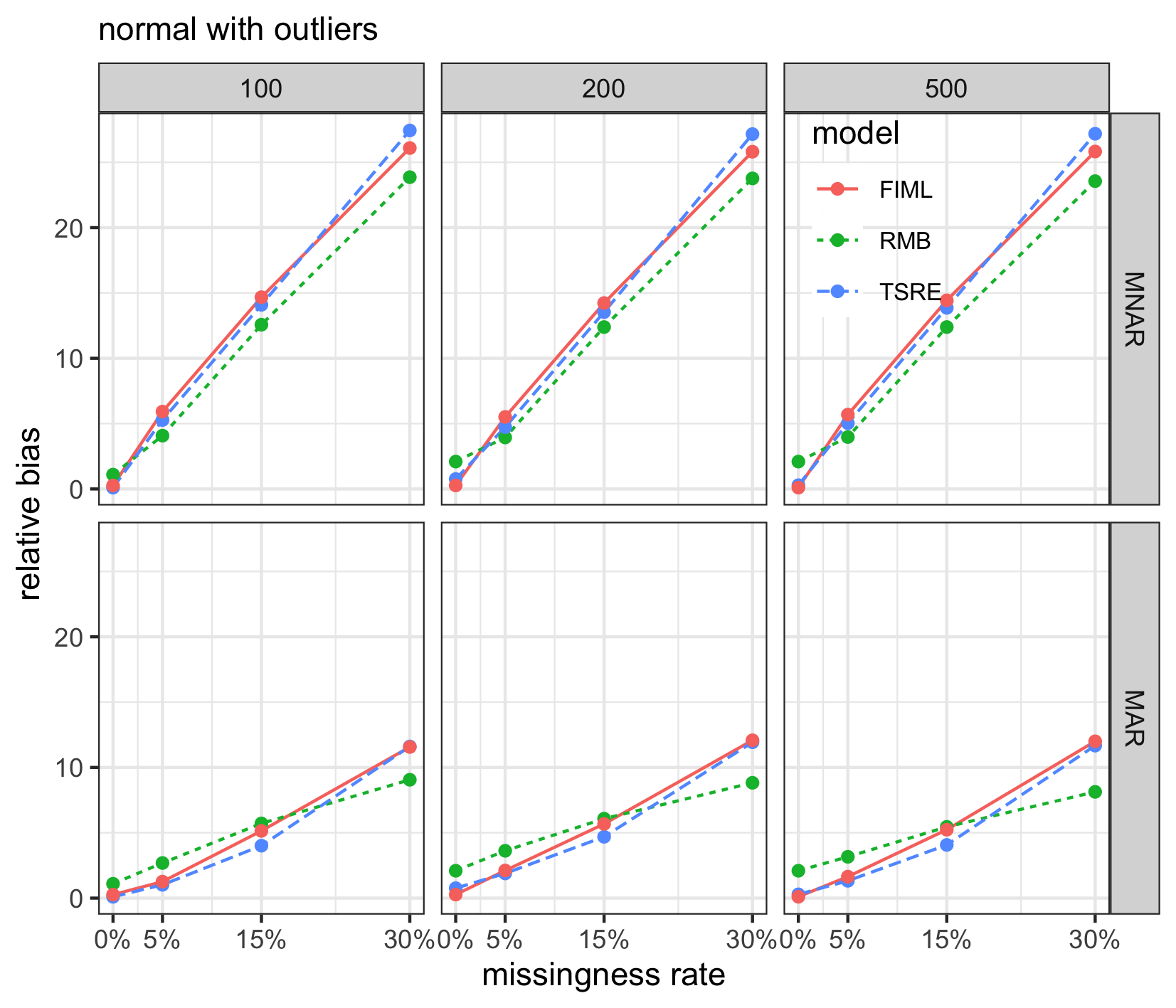} 
    
    \vspace{0.1cm} 
    
    \includegraphics[scale=0.2]{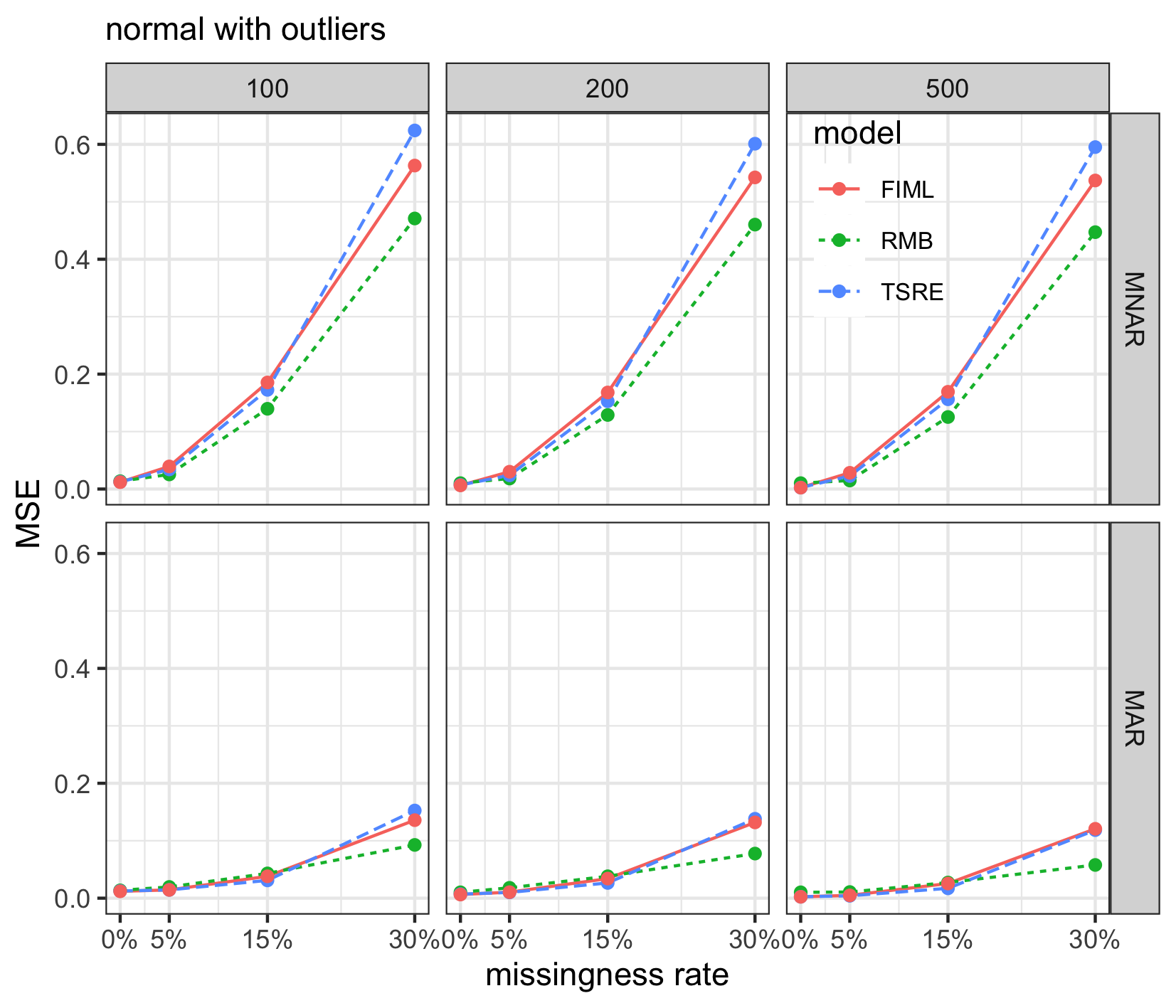} 
    \caption{The relative bias and mean squared error for the average of the latent slope estimates under the normal distribution with outliers} 
    \label{fig:image4}
\end{figure}

\begin{figure}[h]
    \centering
    \includegraphics[scale=0.2]{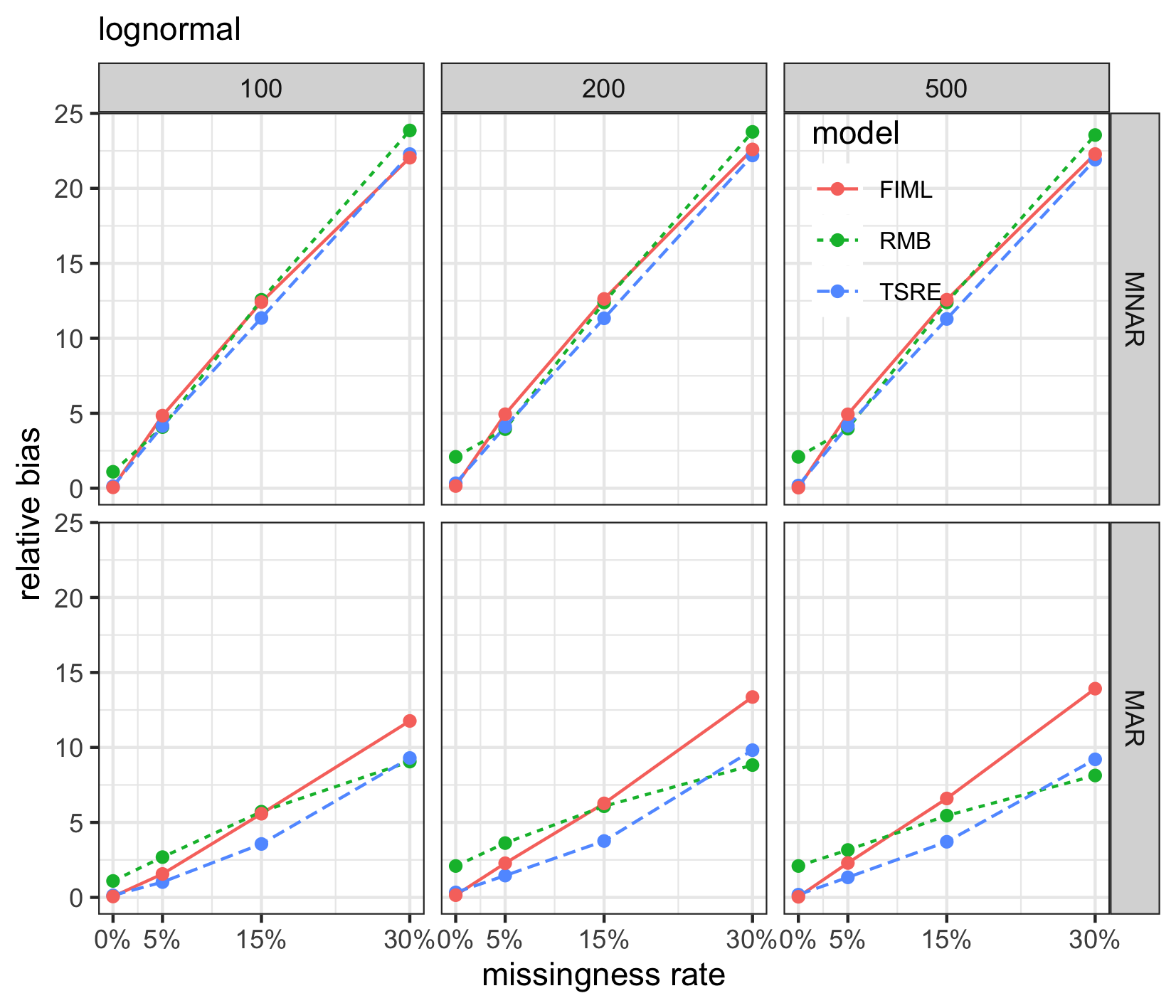} 
    
    \vspace{0.1cm} 
    
    \includegraphics[scale=0.2]{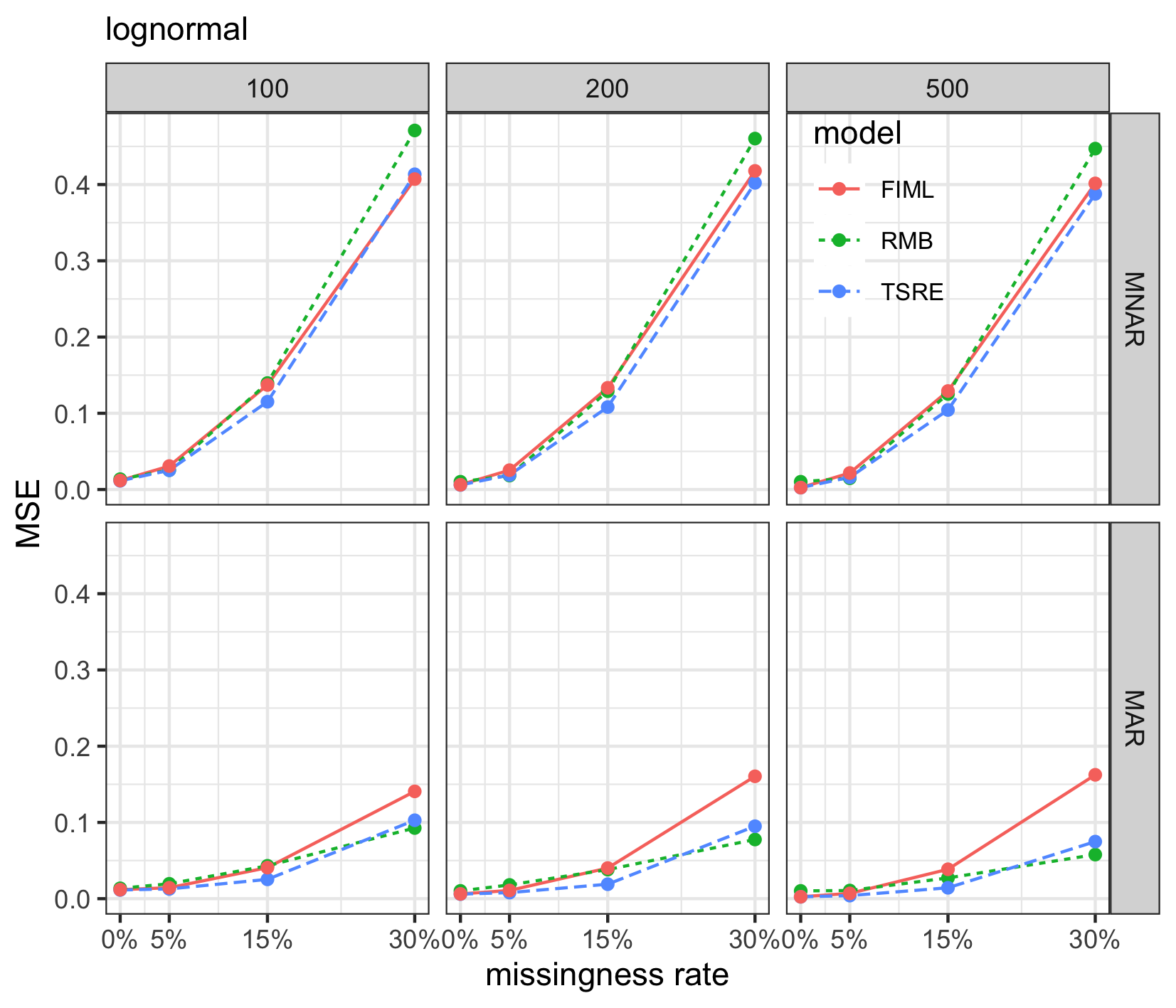} 
    \caption{The relative bias and mean squared error for the average of the latent slope estimates under the lognormal distribution} 
    \label{fig:image5}
\end{figure}

\subsection{Summary}
This simulation study assessed the performances of RMB GCM, FIML, and TSRE under various data conditions. For data following symmetric distributions, RMB GCM, FIML, and TSRE effectively managed the MAR data. For non-symmetrically distributed data, these three methods performed well with MAR data at low to moderate missingness rates; while RMB GCM was the preferred method at higher missingness rates (e.g.,  $>$  15\%). For MNAR data following symmetric distributions, RMB GCM and FIML are preferred approaches, and for data following non-symmetric distributions, RMB GCM was the optimal approach.
\section{An Empirical Example}
After assessing the performances of the three modeling approaches, we illustrate how these approaches can be applied to address real missing data problems.  A real example is provided with a dataset from the National Longitudinal Survey of Youth 1997 Cohort (Bureau of Labor Statistics, U.S. Department of Labor, 2005), where 399 schoolchildren's Peabody Individual Achievement Test (PIAT) math scores were measured yearly from 1997 to 2000.  

Table 2 shows the descriptive statistics of the math scores at each time point, including mean, skewness, kurtosis, bootstrap non-symmetric distribution test result, and missingness rate (ranging from 5.514\% to 12.281 \%).   The bootstrap non-symmetric distribution test (Davison \& Hinkley, 1997) indicates that the PIAT math scores in 1997 are symmetrically distributed, and from 1998 to 2000 are non-symmetrically distributed. Therefore, in general, the observed data could be considered as non-symmetric.
\centerline{[ Insert Table 2 here ]}
\begin{table}[ht]
	\centering
	\caption{Descriptive statistics and symmetrical distribution test}
	\label{t2} 
	\begin{tabular}{rrrrlr}
		\hline
		variable & mean & skewness & kurtosis & symmetrical distribution test & missingness rate\\ 
		\hline
		y1 & 61.16 & -0.08 & 0.21  & -0.73& 6.02\%\\ 
		y2 & 63.27 & -0.66 & 1.01  & -4.26***& 5.51\%\\ 
		y3 & 67.56 & -0.43  & 0.56 & -2.66**& 10.53\%\\ 
        y4  & 69.69  & -0.62 &0.77 & -3.90***& 12.28\%\\  
		\hline
	\end{tabular}
 \begin{tablenotes}
        \footnotesize{*: p \textless .05; **: p \textless .01; ***: p \textless .001.}
      \end{tablenotes}
\end{table} 

FIML, TSRE, and RMB GCM were applied to analyze the data (see the codes on our \href{https://github.com/DandanTang0/GCM-with-Missing-Data}{GitHub} site). Since it is difficult to determine whether the missing data mechanism is MAR or MNAR, we used RMB GCM both with and without the selection structure in Equation (3). We set the total length of the Markov chains to 150,000, with the first 75,000 replications as the burn-in period.  The Geweke convergence test was used to verify the convergence of the chains and suggested that all chains converged.  The parameter estimates of the growth curve model from all these approaches are reported in Table 3.

Based on the simulation results, if the missing data mechanism is MAR,  all these analytical approaches are reliable for this specific dataset because the missing data rate is small (less than 15\%). However, if the missing data mechanism is MNAR,  RMB GCM with the selection structure should be used because the data are non-symmetrically distributed. Thus, in general, we recommended RMB GCM with the selection structure and chose to interpret the model estimation results based on this approach.  The average initial mathematical score in 1997 was approximately 60.60, with a yearly growth rate of 3.17 from 1997 to 2000. The positive estimates of the covariance between the latent intercept and slope suggest that students with higher initial levels of mathematical ability tend to exhibit faster growth rates. Note that the estimated latent slope variance from RMB GCM was very different from those based on FIML and TSRE. This is because in FIML or TSRE, we estimate the between-subject variations in conditional means, while in RMB GCM, we examine the variation in conditional medians, which are more robust against extreme values; the same for the estimated covariance between latent intercept and slope. The estimated latent intercept variance was similar across different estimation methods, because the data at the initial level are relatively symmetric with conditional means very close to conditional medians.

\centerline{[ Insert Table 3 here ]}
\begin{table}[ht]
	\centering
	\caption{Parameter estimates of the growth curve model using RMB GCM, FIML, and TSRE}
	\label{t3}
    \begin{threeparttable}
	\begin{tabular}{rrrrr}
		\hline
       Parameter & RMB GCM& RMB GCM-Selection&FIML& TSRE\\
		\hline
		$\beta_L$  & 60.62* & 60.60* &60.62* & 61.04*\\
		$\beta_S$ & 3.16* & 3.17* &3.10* &3.20*\\
		$\sigma_L^2$  & 181.27* & 181.51* &177.65* & 170.84* \\
         $\sigma_S^2$ & 0.44* & 0.44* &7.31* & 5.83* \\
          $\sigma_{LS}$ & 1.77 & 1.58 &-5.18 & -5.25 \\            
		\hline
	\end{tabular}
 \begin{tablenotes}
        \footnotesize{$\beta_L$: average latent intercepts;
$\beta_S$: average latent slopes;
$\sigma_L^2$: variance of latent intercept;
$\sigma_S^2$: variance of latent slope;
$\sigma_{LS}$: Covariance between the latent intercept and slope. GCM-Selection: GCM with the selection model. 
*: statistical significance.}
      \end{tablenotes}
     \end{threeparttable}
\end{table}

\section{Discussion and Conclusion}
This study used a Monte Carlo simulation to evaluate the performances of three GCM approaches - FIML, TSRE, and RMB, in handling ignorable and nonignorable missing data. Our findings indicate that for symmetric data distributions, FIML, TSRE, and RMB could effectively manage ignorable missing data, while FIML and RMB also excelled in handling nonignorable MNAR data. In scenarios where data did not follow a symmetric distribution, all three approaches were suitable for analyzing MAR data at low to moderate missingness rates, yet RMB GCM emerged as the superior approach under conditions of higher missingness rates (e.g.,  $>$  15\%). Furthermore, RMB GCM outperformed FIML and TSRE for MNAR data in general when the data distribution was not symmetric.

These findings can be understood by noting that, in symmetric data, the mean and median coincide, which explains the similar performance of FIML (based on means) and RMB (based on medians). Under asymmetric conditions, the median typically serves as a more robust measure of central tendency than the mean, giving RMB an advantage over FIML. TSRE can downweight outliers, making it effective when data are contaminated by outliers. However, if the underlying population distribution is systematically skewed, TSRE is not capable to downweight every observation and thus TSRE performs worse than RMB. 

In terms of missing data mechanisms, all three approaches work well for ignorable (MAR) missing data. Yet, when the missing data are nonignorable  (MNAR), FIML cannot account for the missingness process, leading to biased estimation, particularly when the missingness rate is high. TSRE can introduce informative auxiliary variables to transform MNAR data into MAR data (Yuan \& Zhang, 2012). However, identifying appropriate auxiliary variables in practice can be challenging (Curnow et al., 2024; Howard, Rhemtulla, \& Little, 2015). For instance, in our empirical study, no suitable auxiliary variables were available, limiting TSRE’s effectiveness in handling nonignorable missing data.  In addition, TSRE cannot handle nonnormally distributed latent variables (Zhong \& Yuan, 2011), making it not very flexible in terms of its robustness against data distribution.

In contrast, RMB GCM offers greater flexibility and robustness across diverse conditions. First, RMB allows for flexible use of auxiliary variables. However, unlike TSRE, which requires an auxiliary variable, RMB can still function well without one. If an auxiliary variable is unknown, the selection model in Equation (3) can be used; if known, it can be incorporated into the selection model. Second, as demonstrated in our simulation, RMB can effectively handle data with high missingness rates. Third, RMB can potentially handle more complex data types, including categorical data (which TSRE cannot handle), nonnormally distributed data (which FIML struggles with), and even data with nonnormally distributed latent variables.  

When the missing data mechanism is suspected to be MNAR, applying a selection model within RMB GCM is recommended. Numerous selection model forms exist (Enders, 2011; Wu \& Carroll, 1988). We used the model specified in Equation (3) because it is simpler and allows varying missingness probabilities across individually varying time metrics (Tong et al., 2024). If the results from RMB GCM with the selection model are similar to those from RMB GCM without the selection model, missing data may be MAR. Conversely, if the results differ significantly, it indicates the need for a selection model, confirming that the missing data mechanism is MNAR. 
 
In principle, incorporating auxiliary or instrumental variables into the model can convert MNAR data to MAR data, substantially improving the performances of both TSRE and RMB. In the absence of auxiliary variables, TSRE and RMB show reduced performances under higher missingness rates in MNAR scenarios (relative bias > 10\%). Integrating suitable auxiliary variables could mitigate this issue. However, as discussed earlier, identifying such variables in practice is often challenging. In addition, bad instrumental variables or confounders may negatively affect the prediction of missing values and thus result in biased model estimation (Cinelli, Forney, \& Pearl, 2024). Given these complexities, we did not include auxiliary variables in this paper and instead leave this consideration to future research.

Although FIML, TSRE, RMB in growth curve modeling may provide satisfactory estimation results under particular conditions, RMB GCM performs well in general for all data distributions that we investigated and for both ignorable and nonignorable missing data. In practice, if empirical researchers can test that data are symmetrically distributed and have a theory to show the missingness is ignorable, FIML is an easy and standard method to use. However, when substantive researchers are uncertain about the data distribution or the reasons for missingness, or if they believe that missing data may alter the distribution in the sample (e.g., a symmetric population resulting in a nonsymmetric sample), RMB GCM is more reliable to use as it demonstrated greater applicability and flexibility across diverse data conditions than FIML and TSRE.

It is important to note that this study focused on continuous data, while ordinal or categorical data frequently occur in practice. Previous studies suggested that FIML can handle categorical variables by modeling them as observed indicators of underlying latent continuous constructs (Edwards, Berzofsky, \& Biemer, 2015; 2018). TSRE has not yet been developed for use with categorical data. In contrast, RMB offers sufficient flexibility to handle categorical data. Future research should explore whether the findings reported here for continuous data can be generalized to ordinal, categorical, or count data, thereby confirming the robustness and applicability of our conclusions across a broader range of data types.

%\input{reference.bib}
%\renewcommand{\bibsection}{\section*{\flushleft References}}
%\renewcommand{\bibname}{References}
%\bibliographystyle{apalike2}
%\bibliography{reference}

\section{References}
\begin{description}
\item Cinelli, C., Forney, A., \& Pearl, J. (2024). A crash course in good and bad controls. \textit{Sociological Methods \& Research}, \textit{53}(3), 1071-1104.
\item Cruz, K. L. D., Kelsey, C. M., Tong, X., \& Grossmann, T. (2023). Infant and maternal responses to emotional facial expressions: A longitudinal study. \textit{Infant Behavior and Development}, \textit{71}, 101818.
\item Curran, P. J., Obeidat, K., \& Losardo, D. (2010). Twelve frequently asked questions about growth curve modeling. \textit{Journal of cognition and development}, \textit{11}(2), 121-136.
\item Curnow, E., Cornish, R. P., Heron, J. E., Carpenter, J. R., \& Tilling, K. (2024). Multiple imputation using auxiliary imputation variables that only predict missingness can increase bias due to data missing not at random. \textit{BMC medical research methodology}, \textit{24}(1), 231. 
\item Dagdoug, M., Goga, C., \& Haziza, D. (2023). Imputation procedures in surveys using nonparametric and machine learning methods: an empirical comparison. \textit{Journal of Survey Statistics and Methodology}, \textit{11}(1), 141-188.
\item Davison, A. C., \& Hinkley, D. V. (1997). \textit{Bootstrap methods and their application} (No. 1). Cambridge university press. 
\item Depaoli, S., Jia, F., \& Heo, I. (2023). Detecting Model Misspecification in Bayesian Piecewise Growth Models. \textit{Structural Equation Modeling: A Multidisciplinary Journal}, \textit{30}(4), 574-591.
\item Edwards, S. L., Berzofsky, M. E., \& Biemer, P. P. (2015). Addressing nonresponse for categorical data items in complex surveys using full information maximum likelihood. In \textit{Proceedings of the Joint Statistical Meetings, Survey Research Methods Section}. 
\item Edwards, S. L., Berzofsky, M., \& Biemer, P. P. (2018). Addressing nonresponse for categorical data items using full information maximum likelihood with latent GOLD 5.0.
\item Enders, C. K. (2011). Missing not at random models for latent growth curve analyses. \textit{Psychological methods}, \textit{16}(1), 1 
\item Enders, C. K. (2023). Missing data: An update on the state of the art. \textit{Psychological Methods}. 
\item Enders, C. K., \& Bandalos, D. L. (2001). The relative performance of full information maximum likelihood estimation for missing data in structural equation models. \emph{Structural equation modeling, 8(3)}, 430-457.
\item Geweke, J. (1991). \textit{Evaluating the accuracy of sampling-based approaches to the calculation of posterior moments} (No. 148). Federal Reserve Bank of Minneapolis.
\item Grimm, K. J. (2016). \textit{Growth modeling: Structural equation and multilevel modeling approaches}. The Guilford Press.
\item Hoogland, J. J., \& Boomsma, A. (1998). Robustness studies in covariance structure modeling: An overview and a meta-analysis. \textit{Sociological Methods \& Research}, \textit{26}(3), 329-367.
\item Howard, W. J., Rhemtulla, M., \& Little, T. D. (2015). Using principal components as auxiliary variables in missing data estimation. \textit{Multivariate behavioral research}, \textit{50}(3), 285-299.
\item Ibrahim, J. G., \& Molenberghs, G. (2009). Missing data methods in longitudinal studies: a review. \textit{Test}, \textit{18}(1), 1-43.
\item Leeper, J. D., \& Woolson, R. F. (1982). Testing hypotheses for the growth curve model when the data are incomplete. \textit{Journal of Statistical Computation and Simulation}, \textit{15}(2-3), 97-107. 
\item Lei, P. W., \& Shiverdecker, L. K. (2020). Performance of estimators for confirmatory factor analysis of ordinal variables with missing data. \emph{Structural Equation Modeling: A Multidisciplinary Journal, 27(4)}, 584-601.
\item Linero, A. R., \& Daniels, M. J. (2018). Bayesian approaches for missing not at random outcome data: the role of identifying restrictions. \textit{Statistical science: a review journal of the Institute of Mathematical Statistics}, \textit{33}(2), 198.
\item Little, R. J. A., \& Rubin, D. B. (2002). \emph{Statistical analysis with missing data} (2nd ed.). New York, NY: Wiley-Interscience.
\item Micceri, T. (1989). The unicorn, the normal curve, and other improbable creatures. \textit{Psychological bulletin}, \textit{105}(1), 156.  
\item Muthén, L. K., \& Muthén, B. (2017). \emph{Mplus user's guide: Statistical analysis with latent variables}
\item Parodi, K. B., Holt, M. K., Aradhya, P., Green, J. G., \& Merrin, G. J. (2025). A Longitudinal Analysis of Risk and Protective Factors of Bias-Based Bullying Victimization Among Adolescents. \textit{Journal of Interpersonal Violence}, 08862605251318276.
\item Plummer, M. (2003, March). JAGS: A program for analysis of Bayesian graphical models using Gibbs sampling. In \textit{Proceedings of the 3rd international workshop on distributed statistical computing} (Vol. 124, No. 125.10, pp. 1-10).
\item Raykov, T. (2005). Analysis of longitudinal studies with missing data using covariance structure modeling with full-information maximum likelihood. \emph{Structural Equation Modeling, 12}, 493–505.
\item Rosseel, Y. (2012). \emph{lavaan: a brief user's guide}. 
\item Roberts, B. W., Walton, K. E., \& Viechtbauer, W. (2006). Patterns of mean-level change in personality traits across the life course: a meta-analysis of longitudinal studies. \textit{Psychological bulletin}, \textit{132}(1), 1.
\item Schafer, J. L., \& Graham, J. W. (2002). Missing data: our view of the state of the art. 
\item Shi, D., \& Tong, X. (2018). Bayesian robust two-stage causal modeling with nonnormal missing data. \textit{Multivariate Behavioral Research}, \textit{53}(1), 127-127.
\item Shin, T., Davison, M. L., \& Long, J. D. (2017). Maximum likelihood versus multiple imputation for missing data in small longitudinal samples with nonnormality. \textit{Psychological methods}, \textit{22}(3), 426. 
\item Tang, D., \& Tong, X. (2024). Evaluation of Missing Data Analytical Techniques in Longitudinal Research: Traditional and Machine Learning Approaches. \textit{arXiv preprint arXiv:2406.13814}.
\item Tong, X., Kim, S., Bandyopadhyay, D., \& Sun, S. (2024). Association Between Body Fat and Body Mass Index from Incomplete Longitudinal Proportion Data: Findings from the Fels Study. \textit{Journal of Data Science}, \textit{22}(1).
\item Tong, X., Zhang, Z., \& Yuan, K. H. (2014). Evaluation of test statistics for robust structural equation modeling with nonnormal missing data. \textit{Structural Equation Modeling: A Multidisciplinary Journal}, \textit{21}(4), 553-565.
\item Tong, X., Zhang, T., \& Zhou, J. (2021). Robust Bayesian growth curve modelling using conditional medians. \textit{British Journal of Mathematical and Statistical Psychology}, \textit{74}(2), 286-312.
\item Wu, M. C., \& Carroll, R. J. (1988). Estimation and comparison of changes in the presence of informative right censoring by modeling the censoring process. \textit{Biometrics}, 175-188. 
\item Yuan, K. H., \& Bentler, P. M. (2001). Effect of outliers on estimators and tests in covariance structure analysis. \emph{British Journal of Mathematical and Statistical Psychology, 54(1)}, 161-175.
\item Yuan, K. H., Tong, X., \& Zhang, Z. (2015). Bias and efficiency for SEM with missing data and auxiliary variables: Two-stage robust method versus two-stage ML. Structural Equation Modeling: A Multidisciplinary Journal, 22(2), 178-192.
\item Yuan, K. H., \& Zhang, Z. (2011). \textit{rsem: An R package for robust non-normal SEM with missing data}.
\item Yuan, K. H., \& Zhang, Z. (2012). Robust structural equation modeling with missing data and auxiliary variables. \emph{Psychometrika, 77(4)}, 803-826.
\item Zhong, X., \& Yuan, K. H. (2011). Bias and efficiency in structural equation modeling: Maximum likelihood versus robust methods. \textit{Multivariate Behavioral Research}, \textit{46}(2), 229-265.

\end{description}

\end{document}